\documentclass[prl,twocolumn,groupedaddress,showpacs,nofootinbib]{revtex4}
\usepackage{graphicx}
\usepackage{dcolumn}
\usepackage{bm}
\usepackage{amssymb}
\usepackage{amsmath}
\usepackage{epsfig}
\usepackage[dvips]{color}
\usepackage{hhline}
\usepackage{color}

\begin{document}

\title{Signal of Right-Handed Charged Gauge Bosons at the LHC?}

\author{Frank F. Deppisch}
\email{f.deppisch@ucl.ac.uk}
\author{Tomas E. Gonzalo}
\email{tomas.gonzalo.11@ucl.ac.uk}
\affiliation{Department of Physics and Astronomy, University College London, WC1E 6BT, United Kingdom}
\author{Sudhanwa Patra}
\email{sudha.astro@gmail.com}
\affiliation{Centre of Excellence in Theoretical and Mathematical Sciences, Siksha \textquoteleft O\textquoteright 
Anusandhan University, Bhubaneswar, India}
\author{Narendra Sahu}
\email{nsahu@iith.ac.in}
\affiliation{Department of Physics, Indian Institute of Technology, Hyderabad, Yeddumailaram, 502205, Telengana India}
\author{Utpal Sarkar}
\email{utpal@prl.res.in}
\affiliation{Physical Research Laboratory, Ahmedabad 380 009, India}

\begin{abstract}

We point out that the recent excess observed in searches for a right-handed gauge boson $W_R$ at CMS can be explained in a left-right symmetric model with $D$ parity violation. In a class of SO(10) models, in which $D$ parity is broken at a high scale, the left-right gauge symmetry breaking scale is naturally small, and at a few TeV the gauge coupling constants satisfy $g_R \approx 0.6 g_L$. Such models therefore predict a right-handed charged gauge boson $W_R$ in the TeV range with a suppressed gauge coupling as compared to the usually assumed manifest left-right symmetry case $g_R = g_L$. The recent CMS data show excess events which are consistent with the cross section predicted in the $D$ parity breaking model for $1.9~{\rm TeV} < M_{W_R} < 2.4$~TeV. If the excess is confirmed, it would in general be a direct signal of new physics beyond the Standard Model at the LHC. A TeV scale $W_R$ would for example not only rule out SU(5) grand unified theory models. It would also imply $B-L$ violation at the TeV scale, which would be the first evidence for baryon or lepton number violation in nature and it has strong implications on the generation of neutrino masses and the baryon asymmetry in the Universe.

\end{abstract}


\maketitle

\section{Introduction}
One of the popular extensions of the standard model (SM) of particle physics is the left-right symmetric 
model (LRSM) \cite{left-right-group}, which restores parity at higher energies and the observed parity violation 
at low energy is attributed to the different scales of left and right symmetry breaking. At a higher energy 
the SM gauge group ${\cal G}_{SM} \equiv SU(2)_L \times U(1)_{Y}  \times SU(3)_c$ is extended to ${\cal G}_{2213} 
\equiv SU(2)_L \times SU(2)_R \times U(1)_{B-L} \times SU(3)_c$ so that all the left-handed 
particles transform as doublets under $SU(2)_L$ while the right-handed particles transform as doublets under $SU(2)_R$. 
The hypercharge ($Y$) and electric charge ($Q$) are given by
\begin{equation} 
	Q = T_{3L} + T_{3R} + \frac{B-L}{2} = T_{3L} + \frac{Y}{2}.
\end{equation}
Parity then relates the fields transforming under $SU(2)_L$ with that transforming under $SU(2)_R$. The scale of 
$SU(2)_R \times U(1)_{B-L}$ symmetry breaking being different from the scale of electroweak symmetry breaking then 
explains the parity violation at low energy. 

Although a low scale left-right symmetry breaking would provide a very rich phenomenology, it is difficult to justify 
it while being consistent with gauge coupling unification. This problem is solved in the 
$D$ parity breaking LRSM scenarios \cite{D-parity-group}, in which the scalar fields transforming under $SU(2)_L$ can 
have a different mass in comparison to the scalar fields transforming under $SU(2)_R$. As a result, the gauge coupling 
constants $g_L$ and $g_R$ evolve in a different manner and even before the left-right gauge symmetry breaking we find 
$g_L \neq g_R$. In a realistic model, arising from an $SO(10)$ grand unified theory (GUT), we shall show that for a $D$ parity violation at a high 
scale the left-right symmetry can be broken at the TeV scale and gauge coupling unification gives $g_R \approx 0.6 g_L$ before 
the left-right symmetry breaking. This model provides a natural scenario for TeV scale right-handed neutrinos.

Recently, the CMS collaboration at LHC analyzed data to provide a bound on the mass of the right-handed charged gauge 
boson in the LRSM coming from the proton-proton collisions at a center-of-mass energy $\sqrt{s} = 8$~TeV corresponding to 
an integrated luminosity of $19.7~{\rm fb}^{-1}$~\cite{cms-expt}. While the analysis concludes that there is no significant 
discrepancy from SM expectations, the reported data exhibit an intriguing excess of two lepton and two jet events. In this note 
we attempt to interpret this excess in the context of an LRSM with $D$ parity breaking. Considering $g_R (M_{W_R}) \approx 0.6 g_L 
(M_{W_R})$, due to gauge coupling unification, the CMS excess can be interpreted as a signal of right handed charged gauge 
bosons $W_R$ with mass in the range $1.8~{\rm TeV} < M_{W_R} < 2.4~{\rm TeV}$. While the reported excess 
cannot be considered significant we still think it is an interesting hint for what might be the first sign of new physics 
beyond the SM at the LHC.

\section{Left-Right Symmetry with Spontaneous $D$ Parity Breaking}

We start with the matter fields in the model, which transform under $SU(2)_L \times SU(2)_R \times U(1)_{B-L}$ as
\begin{align}
	\ell_{L}=
		\begin{pmatrix}
		  \nu_{L}\\
	    e_{L}
	  \end{pmatrix} \equiv(2,1,-1) &,  \quad
	\ell_{R}=
	\begin{pmatrix} 
		\nu_{R}\\
	  e_{R}
	\end{pmatrix} \equiv(1,2,-1)\,,  \nonumber\\
	Q_{L}=
	  \begin{pmatrix}
	    u_{L} \\
	    d_{L}
	  \end{pmatrix} \equiv(2,1,{\frac{1}{3}}) &,\quad
	Q_{R}=
	  \begin{pmatrix}
	    u_{R}\\
	    d_{R}
	  \end{pmatrix} \equiv(1,2,{\frac{1}{3}})\,.
\end{align}
We further include the standard Higgs fields, belonging to a bidoublet $\Phi$, and two sets of triplet 
fields $\Omega_L$, $\Delta_L$ and $\Omega_R$, $\Delta_R$, to implement the symmetry breaking and provide 
quark and lepton masses, including the neutrino seesaw masses. In addition, we add the singlet scalar 
field $\sigma$ that is odd under $D$ parity breaking it at some high scale without breaking the 
left-right gauge symmetry. These fields transform under ${\cal G}_{LR}$ as
\begin{gather}
	\Delta_{L} \equiv (3,1,-2)\,, \quad
	\Delta_{R} \equiv (1,3,-2)\,, \nonumber \\
	\Omega_{L} \equiv (3,1,0) \,, \quad
	\Omega_{R} \equiv (1,3,0) \,,  \\
	\Phi       \equiv (2,2,0) \,, \quad 
	\sigma     \equiv (1,1,0) \,. \nonumber
\end{gather}
The fermions, Higgs scalars and the vector bosons of the present model transform under the operation of $D$ parity as
\begin{gather}
	\psi_{L,R} \longrightarrow \psi_{R,L}\, , \quad
	\Phi \longrightarrow \Phi^T\, , \quad
	\Delta_{L,R} \longrightarrow \Delta_{R,L}\,, \nonumber \\
	\Omega_{L,R} \longrightarrow  \Omega_{R,L}\, , \quad
	\sigma \longrightarrow -\sigma\, , \quad
	W_{L,R} \longrightarrow W_{R,L}\,.
\label{p-parity}
\end{gather}
The main difference between $D$ parity and parity in the Lorentz group is that the scalar fields $\Delta_{L,R}$ 
and $\Omega_{L,R}$ do not transform under the Lorentz parity while they transform nontrivially under $D$ parity.

$D$ parity is broken when the field $\sigma$ acquires a vacuum expectation value (vev) $\langle \sigma \rangle = \sigma_D$, 
at a scale $M_D$, which is orders of magnitude higher than the left-right gauge symmetry breaking in one or two stages at around the TeV scale,
\begin{equation}
	{\cal G}_{2213} 
	\stackrel{\langle \Omega_R \rangle} {\longrightarrow} {\cal G}_{2113}
	\stackrel{\langle \Delta_R \rangle} {\longrightarrow} {\cal G}_{SM}
\end{equation}
where ${\cal G}_{2113} \equiv SU(2)_L \times U(1)_R \times U(1)_{B-L}
\times SU(3)_c$. $\Omega_{L,R}$ acquires a vev $\langle \Omega_{L,R} \rangle =\omega_{L,R}$ at around $5-10$ TeV yielding a mass of the 
charged right-handed gauge bosons $W_R$ of around $2-4$ TeV. A subsequent breaking step is carried out by $\langle\Delta^0_R\rangle 
\approx v_R$ at around $M_{\rm B-L}\simeq (3-4)$ TeV resulting in a $Z_R$ mass of around $\simeq 1-2$ TeV with present 
experimental bound on $M_{Z_R}$ i.e. $M_{Z_R} \leq 1.162$ TeV \cite{PDG, delAguila:2010mx}. The electroweak symmetry breaking is mediated by 
the usual Higgs doublet $\langle \Phi \rangle =v$. The masses of the heavy states are then given by
\begin{align}
	M^2_{W_R} &\approx g^2_{1R} \omega^2_R, \nonumber \\
	M^2_{Z_R} &\approx \frac{1}{2} 
	\left(g^2_{B-L}+g^2_{1R} \right) (v^2 + 4 v^2_R), \\ 
	M^2_{\Delta_{R/L}} &\approx \mu^2_{\Delta_{R/L}} 
	\pm \lambda \sigma_D M, \nonumber 
\end{align}
where $\lambda$ is a trilinear coupling and the parameters $\langle\sigma\rangle$, $M$, $\mu^2_{\Delta_{R/L}}$ are all ${\cal O}(M_D)$.

We embed this model in a $SO(10)$ grand unified theory, in which the symmetry breaking pattern goes through the Pati--Salam group 
${\cal G}_{224} \equiv SU(2)_L \times SU(2)_R \times SU(4)_c$ as~\cite{D-parity-group,sudhanwa_work}
\begin{align}
\label{eq:BreakingChain}
	SO(10) &\mathop{\longrightarrow}^{M_U} \mathcal{G}_{224D} \,
  	     \mathop{\longrightarrow}^{M_D}_{} \mathcal{G}_{224} \,
   	    \mathop{\longrightarrow}^{M_C}_{} \mathcal{G}_{2213} \, 
				\nonumber \\
   	    &\mathop{\longrightarrow}^{M_{\Omega}} \mathcal{G}_{2113}\,
   	    \mathop{\longrightarrow}^{M_{B-L}} \mathcal{G}_{SM}\,
   	    \mathop{\longrightarrow}^{M_{Z}} \mathcal{G}_{13}\, . 
\end{align}
\begin{figure}[t]
\centering
\includegraphics[width=0.95\linewidth]{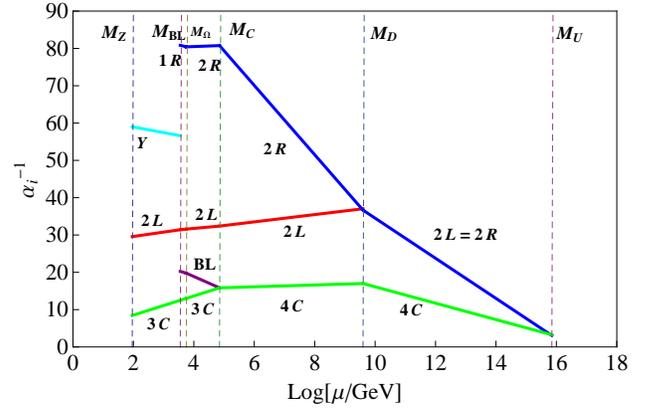}
\caption{One loop renormalization group evolution of gauge couplings with Pati--Salam symmetry $G_{224D}$ as the highest  
        intermediate symmetry breaking where the inverse fine structure constant $\alpha^{-1}_i$ is plotted against energy
        scale with $i=Y, 2L, 1R, 2R, 3C, 4C, B-L$ defined at appropriate scales. The numerical values for different 
        intermediate mass scales, denoted at the top edge of the plot, are presented in the text.}
\label{fig:RG-evolution}
\end{figure}

The symmetry breaking of $SO(10)$ to the SM is achieved by the Higgs multiplets $10_{H}$, $126_{H}$, $54_{H}$ and $210_{H}$. However, 
we have introduced two extra Higgs multiplets $16_{H}$ and $210_{H}$ in the renormalization group evolution to achieve the unification 
of gauge couplings. This is shown in Fig.~\ref{fig:RG-evolution}. From the gauge coupling unification, the intermediate 
mass scales are found to be $M_{B-L}=(3-4)$~TeV, $M_{\Omega}=5-10$~TeV, $M_{C}=10^{5}-10^{6}$~GeV, $M_D=10^{9.6}$~GeV and $M_U=10^{15.89}$~GeV. 
The most desirable prediction of the model is that the values of $g_L$ and $g_R$ at TeV scale, consistent with gauge coupling 
unification, are given by $g_L \approx 0.632$ and $g_R \approx 0.38$. As a result the ratio of right- and left-handed $SU(2)$ gauge 
couplings around the TeV scale is found to be
\begin{align}
	\frac{g_R}{g_L} \approx 0.6\,.
\end{align}

The model presented here is of course not the only possibility to achieve nonuniversal left and right gauge couplings in the desired range. For example, a survey of models based on SO(10) without manifest left-right symmetry is presented in Ref.~\cite{romao_paper}.

\section{Right-handed $W$ Boson and Neutrino Searches at the LHC}

It is interesting that this particular ratio of the gauge coupling strengths allows us to interpret the excess of events at CMS \cite{cms-expt} as the signature of a right-handed charged gauge boson. The CMS collaboration looked for a signature with two leptons and two jets arising 
from the resonant production of a $W_R$ boson, which decays through a right-handed neutrino $N$ as \cite{Keung:1983uu}
\begin{align}
\label{eq:lhcprocess}
	p p \to W_R \to l_1 N \to l_1 l_2 W_R^* \to l_1 l_2 + 2\text{ jets}.
\end{align}
As shown, the heavy neutrino decays through an off-shell $W_R$. Other decays of a $W_R$ have been discussed in \cite{Torre_2011}. The CMS analysis treats events with two electrons and two muons separately. On the other hand, it does not differentiate between different lepton charges, but among the 14 potential signal events only one same sign lepton event was seen. As the heavy neutrino is considered a Majorana fermion 
in the LRSM, both opposite sign ($l_1^\pm l_2^\mp$) and same sign events ($l_1^\pm l_2^\pm$) are expected. To explain this discrepancy, nonminimal seesaw sectors incorporating quasi-Dirac heavy neutrinos have to be employed. Such a scenario is for example generally expected for TeV scale heavy neutrinos and large Yukawa couplings. The CMS analysis does not consider 
possible lepton flavour violating signatures containing both an electron and a muon. Using samples collected at a center-of-mass energy of 8~TeV with 
an integrated luminosity of $19.7$~fb$^{-1}$, no significant excess is reported and $W_R$ masses $m_{W_R} < 2.87$ (3.00)~TeV are excluded in the 
$ee$ ($\mu\mu$) channel, for $M_N = \frac{1}{2} M_{W_R}$ (this corresponds to the approximate extent of the excluded $M_{W_R} - M_N$ 
parameter space). Intriguingly, the collected data exhibit an excess in the $ee$ channel with a local significance of $2.8\sigma$ for the 
invariant and candidate $W_R$ mass $M_{eejj} = M_{W_R} \approx 2.1$~TeV. No excess is seen in the $\mu\mu$ channel. The CMS analysis also does not see any localized excess in the distribution expected from the decay $N \to l_2 + 2\text{ jets}$ in the signal process \eqref{eq:lhcprocess}, but it is not obvious how this would affect the significance of ruling out this specific process. For example, the CMS analysis simulated events for scenarios with $M_N = \frac{1}{2} M_{W_R}$, and it is not clear how the stated conclusion that no other localized excesses are seen applies to the general case $M_N \neq \frac{1}{2} M_{W_R}$ as well. 

\begin{figure}[t]
\centering
\includegraphics[width=0.9\linewidth]{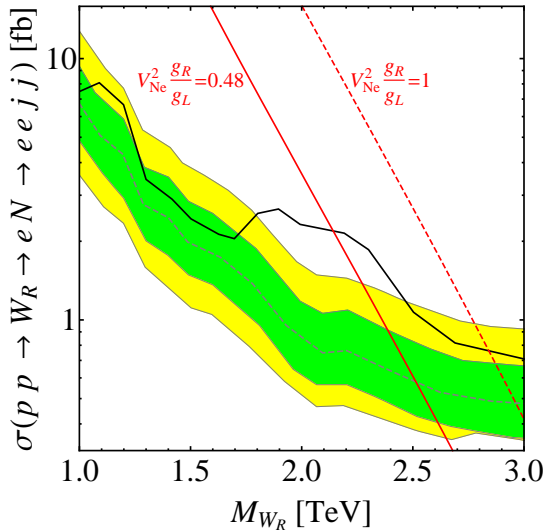}
\caption{Theoretically predicted cross sections (red curves) and experimental exclusion limits as a function of the $W_R$ mass. The dashed red curve corresponds to the standard scenario $g_R = g_L$, $V_{Ne} = 1$ and the solid red curve to the $D$ parity breaking scenario $g_R/g_L = 0.6$, $V_{Ne} = 0.9$ discussed in the text. The observed (solid black curve) and expected (dashed grey curve and green / yellow bands) 95\% exclusion limits are taken from Ref.~\cite{cms-expt}.}
\label{fig:W_R-excess}
\end{figure}

The CMS analysis compares the experimental result 
with the theoretically predicted cross section in the minimal LRSM using $g_L = g_R$. In this case, the $ee$ excess cannot be explained by the theoretical prediction since the predicted cross section is too large by a factor of $\approx 3-4$. This discrepancy could be reconciled in our model with a smaller $g_R$. In the following we assume that the excess is due to the 
production of a $W_R$ which decays to a heavy neutrino $N$ that dominantly couples to electrons with a large right-handed current mixing 
matrix element $V_{Ne} \lesssim 1$. This can for example be achieved with a normally ordered hierarchical spectrum of three heavy neutrinos with small mixing between the generations. We also assume that there is a negligible mixing between the heavy and light neutrinos as well as the left and right $W$ bosons\footnote{Sizable mixing between left and right neutrinos or $W$ bosons could be an alternative explanation for the reduced cross section as the decay branching ratios in \eqref{eq:lhccs} will be reduced; see for example Ref.~\cite{Chen:2013foz}.}. In this case, both $W_R$ and $N$ couple only through right-handed currents and the total 
cross section of the process under consideration can be expressed as
\begin{align}
\label{eq:lhccs}
	\sigma (pp \to eejj) &= \sigma (pp \to W_R) \nonumber\\
	&\quad\times {\rm Br} (W_R \to e N) 
	      \times {\rm Br} (N\to ejj) \nonumber\\
	&= V_{Ne}^4 \left(\frac{g_R}{g_L}\right)^2 \sigma_\text{CMS}(pp \to eejj),
\end{align}
where $\sigma_\text{CMS}(pp \to eejj)$ corresponds to the scenario with $g_L = g_R$ and $V_{Ne} = 1$ as used in the CMS analysis. Instead, using 
the value derived in the LRSM with spontaneous symmetry breaking and SO(10) unification, the predicted cross section is suppressed by a factor of 
$\approx 0.4$. This is already sufficient to allow the excess to be interpreted as a signal. In addition, even a small deviation in $V_{Ne}\approx 
0.9$ will lead to a sizable further suppression. This is shown in Fig.~\ref{fig:W_R-excess} where we compare our calculated process cross section 
with the CMS result. The dashed red curve gives the predicted cross section as a function of $M_{W_R}$ and $M_N = \frac{1}{2} M_{W_R}$ for $g_R = g_L$ and $V_{Ne} = 1$ (essentially coinciding with the corresponding curve in the CMS analysis), whereas the solid red curve corresponds to $g_R/g_L = 0.6$ and $V_{Ne} = 0.9$. The solid black curve is the observed CMS 95\% exclusion whereas the dashed grey curve and green (yellow) bands show the expected 95\% exclusion with $1\sigma$ ($2\sigma$) uncertainty, with an excess in the region $1.9 \text{ TeV} \lesssim M_{W_R} \lesssim 2.4$~TeV. It is clear that the simple modification arising in the LRSM with $D$ parity breaking can successfully explain the observed excess.
In our scenario, the suppression of the cross section is mainly due to the smaller gauge coupling but also due to $V_{Ne} = 0.9$. This choice was simply made to emphasize that (part of) the reduction in a given channel ($ee$, $\mu\mu$, $\tau\tau$) could simply arise from large intergenerational mixing. Because of unitarity, $|V_{Ne}|^2 + |V_{N\mu}|^2 +|V_{N\tau}|^2 = 1$ (assuming negligible left-right mixing), large deviations from unity, $V_{Ne} = 1$, would also produce $\mu\mu$ and/or $\tau\tau$ events, whereas a smaller value for $g_R$ suppresses all channels. In addition, large mixing among the right-handed neutrinos should be avoided, especially in the $e\mu$ sector and for strongly hierarchical heavy neutrinos, as it induces large lepton flavour violation.

It is of course important to take into account other constraints on the $W_R$ mass as well. Previous searches for $W_R \to l N$ at CMS \cite{CMS:2012zv} and ATLAS \cite{ATLAS:2012ak} did not see any excess and report an exclusion at 95\% confidence level of $M_{W_R} \gtrsim 2.5$~TeV. While apparently incompatible with a signal at $\approx 2.1$~TeV, this limit would also have to be adjusted in our $D$ parity scenario using $V_{Ne}^2 \frac{g_R}{g_L} \approx 0.5$. We estimate that the previous LHC limit would weaken to $M_{W_R} \gtrsim 2.1$~TeV. A similar argument would also apply to the already weaker direct limits from $W_R \to t \bar b$ decay searches at the LHC \cite{OtherLhcSearches}. The strongest indirect bound on $M_{W_R}$ is due to the $K_L - K_S$ mass difference~\cite{KLKS},
\begin{equation}
	|h_K| \approx \left(\frac{g_R}{g_L}\right)^2 
	              \left(\frac{2.4\text{ TeV}}{M_{W_R}}\right)^2
				< 1.
\end{equation}
In the standard scenario, this leads to the bound $M_{W_R} \gtrsim 2.5$~TeV, whereas for $\frac{g_R}{g_L} = 0.6$ the limit weakens to $M_{W_R} \gtrsim 1.5$~TeV, compatible with the potential signal at $M_{W_R} \approx 2.1$~TeV.

\section{Conclusion}
We have shown that a TeV scale left-right symmetric model can naturally arise via spontaneous $D$ parity breaking. The asymmetry between the gauge 
couplings near the electroweak symmetry breaking scale is then a consequence of gauge coupling unification. Assuming that the Pati--Salam symmetry $SU(2)_L \times SU(2)_R \times SU(4)_C$ is the largest subgroup of a nonsupersymmetric SO(10) grand unified theory we obtain $g_R/g_L \approx 0.6$. This gives rise to an extra suppression in the production of $W_R$ in proton-proton collisions. As a result we could reconcile our prediction for $W_R \to eejj$ events at the LHC with the recent $2.8 \sigma $ excess within the mass 
range $1.9~{\rm TeV} < M_{W_R} < 2.4$~TeV, reported recently by the CMS collaboration. If this result is confirmed by future data, {\sl it would be the first direct evidence for physics beyond the standard model from the LHC, which would
rule out the SU(5) GUT. Moreover, a TeV scale $W_R$ would imply $B-L$ violation at the TeV scale (which would also be the first evidence for baryon or lepton number violation), which has strong implication on the generation of baryon asymmetry of the Universe as well as the mechanism of neutrino mass generation}. For example, if the excess were to be confirmed for the same sign lepton events, sizable contributions to neutrinoless double beta decay are possible and high scale models of leptogenesis would be strongly disfavoured \cite{Deppisch:2013jxa}. While the excess cannot be considered a significant deviation from the SM as of now, the model we discussed here demonstrates that the excess can be explained in well-motivated extensions of the minimal left-right symmetric model. In light of the theoretical importance we therefore suggest to put a focus on further studies of the excess. In addition to possible enhancements of the excess significance by focusing on the kinematic region of interest, this could 
include analyses of the presence of lepton number and potential lepton flavour violating \cite{Das:2012ii} components in the excess.

\section*{Acknowledgments}
The work of FFD and TEG was supported partly by the London Centre for Terauniverse Studies (LCTS), using funding from the European Research Council via the Advanced Investigator Grant 267352. FFD would like to thank Abhiraami Navaneethanathan for useful discussions and for the help in compiling the analysis data. The work of SP is supported by the Department of Science and Technology, Government of India under the financial Grant SERB/F/482/2014-15. NS is partially supported by the Department of Science and Technology Grant SR/FTP/PS-209/2011. The work of US is partially supported by the J.C. Bose National Fellowship grant from the Department of Science and Technology, India.

\end{document}